\documentclass[aps,nofootinbib]{revtex4}
\usepackage{graphicx}
\usepackage{latexsym}
\usepackage{revsymb}
\usepackage{amsfonts}
\usepackage{amsmath}
\usepackage{amssymb}
\usepackage{bm}

\newcommand{\ket}[1]{| #1 \rangle}

\newcommand{\vev}[1]{\left\langle #1 \right\rangle}
\newcommand{\bvev}[1]{\bigl\langle #1 \bigr\rangle}

\newcommand{\ellP}{\ell_\mathrm{P}}
\newcommand{\muP}{\mu_\mathrm{P}}

\newcommand{\muSUSY}{\mu_\mathrm{SUSY}}
\newcommand{\GN}{G_\mathrm{N}}
\newcommand{\Ahat}{\hat{A}}
\newcommand{\Bhat}{\hat{B}}
\newcommand{\Hhat}{\hat{H}}
\newcommand{\Ohat}{\hat{O}}
\newcommand{\xhat}{\hat{x}}

\newcommand{\phat}{\hat{p}}
\newcommand{\qhat}{\hat{q}}
\newcommand{\rhat}{\hat{r}}
\newcommand{\vhat}{\hat{v}}
\newcommand{\pbhat}{\hat{\mathbf{p}}}
\newcommand{\qbhat}{\hat{\mathbf{q}}}
\newcommand{\xbhat}{\hat{\mathbf{x}}}
\newcommand{\pb}{\mathbf{p}}

\newcommand{\xb}{\mathbf{x}}
\newcommand{\Ai}{\mathrm{Ai}}

\begin{document}

\title{On the Minimal Length Uncertainty Relation and the Foundations of String Theory}

\author{Lay Nam Chang}\email{laynam@vt.edu}
\author{Zachary Lewis}\email{zlewis@vt.edu}
\author{Djordje Minic}\email{dminic@vt.edu}
\author{Tatsu Takeuchi}\email{takeuchi@vt.edu}

\affiliation{Department of Physics, Virginia Tech, Blacksburg, VA 24061, USA}
\date{May 31, 2011}

\begin{abstract}
We review our work on the minimal length
uncertainty relation as suggested by perturbative
string theory. We discuss simple phenomenological
implications of the minimal length uncertainty
relation and then argue that the combination of the
principles of quantum theory and general relativity
allow for a dynamical energy-momentum space.
We discuss the implication of this for the problem of
vacuum energy and the foundations of non-perturbative string theory.
\end{abstract}

\maketitle
\section{Introduction}

One of the unequivocal characteristics of string theory \cite{string-textbooks} is its
possession of a fundamental length scale which determines the typical spacetime extension of a fundamental string.
This is $\ell_s = \sqrt{\alpha'}$, where $\hbar c/\alpha'$ is the string tension.
Such a feature is to be expected of any candidate theory of quantum gravity,
since gravity itself is characterized by the Planck length $\ellP = \sqrt{\hbar\GN/c^3}$.
Moreover, $\ellP\sim\ell_s$ is understood to be the 
\textit{minimal length} below which spacetime distances cannot be resolved \cite{Wheeler:1957mu,MinimalLength}:
\begin{equation}
\delta s \;\agt\; \ellP \sim \ell_s\;.
\label{MinimalLength}
\end{equation}
Quantum theory, on the other hand, 
is completely oblivious to the presence of such a scale,
despite its being the putative infrared limit of string theory.
A natural question to ask is, therefore, whether the formalism of quantum theory can be deformed or
extended in such a way as to consistently incorporate the minimal length.  
If it is at all possible, the precise manner in which quantum theory must be modified
may point to solutions of yet unresolved mysteries such as the cosmological constant problem \cite{CosmoConstant},
which is quantum gravitational in its origin. 
It should also illuminate the nature of string theory \cite{joewhat}, whence quantum theory must emerge \cite{CLMTT}.

The idea of introducing a minimal length into quantum theory has a fascinating and long history.
It was used by Heisenberg in 1930 \cite{kragh} to address
the infinities of the newly formulated theory of quantum electodynamics \cite{Heisenberg:1929xj}.
Over the years, the idea has been picked up by many authors in a plethora of contexts, e.g. 
Refs.~\cite{born:1933,Snyder:1946qz,Yang:1947ud,Mead:1966zz,Pavlopoulos:1967dm,Padmanabhan:1986ny,Hossenfelder:2003jz,Das:2008kaa,Bagchi:2009wb} to list just a few.
Various ways to deform or extend quantum theory have also been suggested \cite{Weinberg:1989cm,Bender:2002vv}.
In this paper, we focus our attention on how a minimal length can be introduced into quantum mechanics
by modifing its algebraic structure \cite{Maggiore:1993zu,Kempf:1994su}.


The starting point of our analysis is the minimal length
uncertainty relation (MLUR) \cite{Amati:1988tn},
\begin{equation}
\delta x \;\sim\;
\left(\dfrac{\hbar}{\delta p} + \alpha'\,\dfrac{\delta p}{\hbar}\right)\;,
\label{MLUR}
\end{equation}
which is suggested by a re-summed
perturbation expansion of the string-string scattering amplitude in a flat spacetime background \cite{gross}.
This is essentially a Heisenberg microscope argument \cite{HeisenbergMicroscope}
in the S-matrix language \cite{Smatrix} with fundamental strings used to probe fundamental strings.
The first term inside the parentheses on the right-hand side is the usual
Heisenberg term coming from the shortening of the probe-wavelength as momentum is increased,
while the second-term can be understood as due to the lengthening of the probe string as more energy is pumped into it:
\begin{equation}
\delta p 
\;=\; \dfrac{\delta E}{c} 
\;\sim\; \dfrac{\hbar}{\alpha'}\,\delta x 
\;.
\label{dPproptodX}
\end{equation}
Eq.~(\ref{MLUR}) implies that the uncertainty in position, $\delta x$, is bounded from 
below by the string length scale,
\begin{equation}
\delta x \;\agt\; \sqrt{\alpha'} \;=\; \ell_s \;, 
\label{dXmin}
\end{equation}
where the minimum occurs at 
\begin{equation}
\delta p \;\sim\; \dfrac{\hbar}{\sqrt{\alpha'}} \;=\; \dfrac{\hbar}{\ell_s} \;\equiv\; \mu_s\;.
\end{equation}
Thus, $\ell_s$ is the minimal length below which spatial distances cannot be resolved,
consistent with Eq.~(\ref{MinimalLength}).
In fact, the MLUR can be motivated by fairly elementary general relativistic considerations 
independent of string theory, which suggests that it is a universal feature of quantum gravity \cite{Wheeler:1957mu,MinimalLength}.

Note that in the trans-Planckian momentum region $\delta p \gg \mu_s$, the MLUR is
dominated by the behavior of Eq.~(\ref{dPproptodX}), which implies that large $\delta p$ (UV)
corresponds to large $\delta x$ (IR), and that there exists a correspondence
between UV and IR physics.
Such UV/IR relations have been observed in various string dualities \cite{string-textbooks},
and in the context of AdS/CFT correspondence \cite{uvir}
(albeit between the bulk and boundary theories).
Thus, the MLUR captures another distinguishing feature of string theory.

In addition to the MLUR, 
another uncertainty relation has been pointed out by Yoneya as characteristic of string theory.
This is the so-called spacetime uncertainty relation (STUR)
\begin{equation}
\delta x\,\delta t \;\sim\; \ell_s^2/c\;,
\label{STUR}
\end{equation}
which can be motivated in a somewhat hand-waving manner by combining the usual energy-time uncertainty relation $\delta E\,\delta t\sim \hbar$ \cite{energytimeUR} with Eq.~(\ref{dPproptodX}). 
However, it can also be supported via an analysis of D0-brane scattering in certain backgrounds 
in which $\delta x$ can be made arbitrary small 
at the expense of making the duration of the interaction $\delta t$ arbitrary large \cite{yoneya}.
While the MLUR pertains to dynamics of a particle in a non-dynamic spacetime, the STUR
can be interpreted to pertain to the dynamics of spacetime itself in which 
the size of a quantized spacetime cell is preserved.

In the following, we discuss how the MLUR and STUR may be 
incorporated into quantum mechanics via a deformation and/or extension of its algebraic structure.
In section~II, we introduce a deformation of the canonical commutation relation between
$\xhat$ and $\phat$ which leads to the MLUR,
and discuss its phenomenological consequences.
In section~III, we take the classical limit by replacing commutation relations
with Poisson brackets and derive the analogue of Liouville's theorem in the deformed
mechanics. We then discuss the effect this has on the density of states in phase space.
In section~IV, we discuss the implications of the MLUR on the cosmological constant problem.
We conclude in section~V with some speculations on how the STUR may be
incorporated via a Nambu triple bracket,
and comment on the lessons for the foundations of string theory and on the question ``What is string theory?''

\section{Quantum Mechanical Model of the Minimal Length}

\subsection{Deformed Commutation Relations}

\begin{figure}[ht]
\includegraphics[width=8cm]{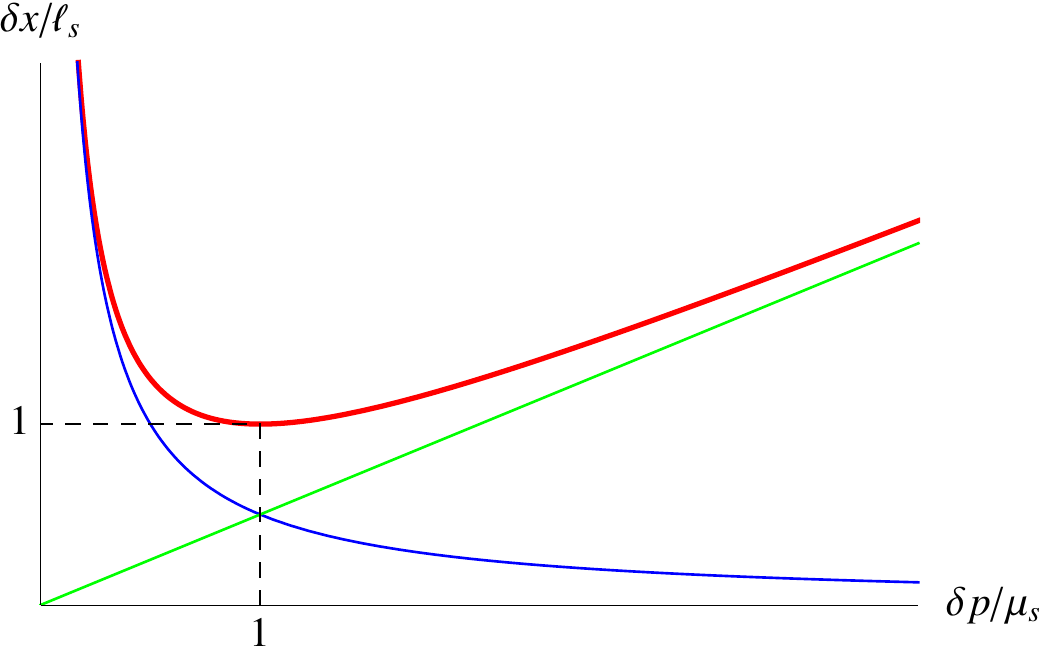}
\caption{The $\delta p$-dependence of the lower bound of $\delta x$
under the minimal length uncertainty relation Eq.~(\ref{MLUR2}) (red curve).
The bound for the usual Heisenberg relation $\delta x\ge \hbar/(2\delta p)$ is shown in blue,
and the linear bound $\delta x\ge (\hbar\beta/2)\delta p$ is shown in green.}
\label{MLURfig}
\end{figure}

To place the MLUR, Eq.~(\ref{MLUR}), on firmer ground, we begin by rewriting it as
\begin{equation}
\delta x\,\delta p \;\ge\; \dfrac{\hbar}{2}\left(1 + \beta\, {\delta p}^2\right)\;,
\label{MLUR2}
\end{equation}
where we have introduced the parameter $\beta = \alpha'/\hbar^2$.
The minimum value of $\delta x$ as a function of $\delta p$ is plotted in Fig.~\ref{MLURfig}.
This uncertainty relation can be reproduced by deforming the canonical commutation 
relation between $\xhat$ and $\phat$ to:
\begin{equation}
\dfrac{1}{i\hbar}\left[\,\xhat,\,\phat\,\right]\;=\; 1
\qquad\longrightarrow\qquad
\dfrac{1}{i\hbar}\left[\,\xhat,\,\phat\,\right]\;=\; A(\phat^2)\;,
\label{MLUR3}
\end{equation}
with $A(p^2)=1+\beta p^2$. Indeed, we find
\begin{equation}
\delta x\,\delta p\;\ge\;
\dfrac{1}{2}\Bigl|\bvev{\left[\,\xhat,\,\phat\,\right]}\Bigr|
\;=\; \dfrac{\hbar}{2}
\left(1+\beta\vev{\phat^2}\right)
\;\ge\; \dfrac{\hbar}{2}
\left(1+\beta\,\delta p^2\right)\;,
\end{equation}
since $\delta p^2 = \vev{\phat^2}-\vev{\phat}^2$.
The function $A(p^2)$ can actually be more generic, with $\beta p^2$ being the
linear term in its expansion in $p^2$.

When we have more than one spatial dimension, the above commutation relation can
be generalized to
\begin{equation}
\dfrac{1}{i\hbar}[\,\xhat_i,\,\phat_j\,]\;=\; A(\pbhat^2)\,\delta_{ij}+B(\pbhat^2)\,\phat_i\phat_j\;,
\label{deformedXP}
\end{equation}
where $\pbhat^2 = \sum_i \phat_i^2$.
The right-hand side is the most general form that depends only on the momentum and 
respects rotational symmetry.  
Assuming that the components of the momentum commute among themselves,
\begin{equation}
[\,\phat_i,\,\phat_j\,] \;=\; 0\;,
\label{PP}
\end{equation}
the Jacobi identity demands that
\begin{equation}
\dfrac{1}{i\hbar}[\,\xhat_i,\,\xhat_j\,]
\;=\; -\left\{\,2(\Ahat+\Bhat\pbhat^2)\Ahat' - \Ahat\Bhat\,\right\} \hat{L}_{ij}\;,
\label{deformedXX}
\end{equation}
where we have used the shorthand $\Ahat=A(\pbhat^2)$, $\Ahat'=\dfrac{dA}{d\pb^2}(\pbhat^2)$, $\Bhat=B(\pbhat^2)$, 
and $\hat{L}_{ij} = \left(\xhat_i\phat_j-\xhat_j\phat_i\right)/\Ahat$
. That $\hat{L}_{ij}$ generates rotations can be seen from the following:  
\begin{eqnarray}
\dfrac{1}{i\hbar}[\,\hat{L}_{ij}\,\xhat_{k}\,]
& = & \delta_{ik}\xhat_j - \delta_{jk}\xhat_i \;,
\cr
\dfrac{1}{i\hbar}[\,\hat{L}_{ij}\,\phat_{k}\,]
& = & \delta_{ik}\phat_j - \delta_{jk}\phat_i \;,
\cr
\dfrac{1}{i\hbar}[\,\hat{L}_{ij}\,\hat{L}_{k\ell}\,]
& = & \delta_{ik}\hat{L}_{j\ell} - \delta_{i\ell}\hat{L}_{jk}
    + \delta_{j\ell}\hat{L}_{ik} - \delta_{jk}\hat{L}_{i\ell}
\;.
\end{eqnarray}
Note that the non-commutativity of the components of position
can be interpreted as a reflection of the dynamic nature of space itself,
as would be expected in quantum gravity.

Various choices for the functions $A(\pb^2)$ and $B(\pb^2)$ have been considered in the literature.
Maggiore \cite{Maggiore:1993zu} proposed
\begin{equation}
A(\pb^2)\;=\;\sqrt{1+2\beta\pb^2}\;,\qquad 
B(\pb^2)\;=\;0\;,\qquad
\dfrac{1}{i\hbar}[\,\xhat_i,\,\xhat_j\,]\;=\; -2\beta \hat{L}_{ij}\;,
\label{MaggioreAB}
\end{equation}
while Kempf \cite{Kempf:1994su} assumed 
\begin{equation}
A(\pb^2)\;=\;1+\beta \pb^2\;,\qquad B(\pb^2)\;=\;\beta'\;=\;\mathrm{constant}\;,
\label{KempfAB}
\end{equation}
in which case
\begin{equation}
\dfrac{1}{i\hbar}[\,\xhat_i,\,\xhat_j\,]
\;=\; -\left\{\,(2\beta-\beta')+\beta(2\beta+\beta')\pbhat^2\right\} \hat{L}_{ij}\;.
\end{equation}
Kempf's choice encompasses the algebra of Snyder \cite{Snyder:1946qz}
\begin{equation}
A(\pb^2)\;=\;1\;,\qquad 
B(\pb^2)\;=\;\beta'\;,\qquad
\dfrac{1}{i\hbar}[\,\xhat_i,\,\xhat_j\,]\;=\; \beta'\hat{L}_{ij}\;,
\label{SnyderAB}
\end{equation}
and that of Brau \cite{Brau:1999uv,Brau:2006ca}
\begin{equation}
A(\pb^2)\;=\;1+\beta\pb^2\;,\qquad 
B(\pb^2)\;=\;2\beta\;,\qquad
\dfrac{1}{i\hbar}[\,\xhat_i,\,\xhat_j\,]\;=\; O(\beta^2)\;,
\label{BrauAB}
\end{equation}
for which the components of the position approximately commute.
In our treatment, we follow Kempf and use Eq.~(\ref{KempfAB}).
%

\subsection{Shifts in the Energy Levels}

Let us see whether the above deformed commutation relations lead to
a reasonable quantum mechanics, with well defined energy eigenvalues 
and eigenstates.
Given a Hamiltonian in terms of the deformed position and momentum operators, 
$H(\xbhat,\pbhat)$, we would like to solve the time-independent Schr\"odinger equation
\begin{equation}
H(\xbhat,\pbhat)\,\ket{E} \;=\; E\,\ket{E}\;.
\end{equation}
The operators which satisfy 
Eqs.~(\ref{deformedXP}), (\ref{PP}), and (\ref{deformedXX}), subject to
the choice Eq.~(\ref{KempfAB}), can be represented 
using operators which obey the canonical commutation relation 
$[\,\qhat_i,\,\phat_j\,]=i\hbar\,\delta_{ij}$ as \cite{Kempf:1994su,Benczik:2007we}
\begin{eqnarray}
\xhat_i & = & \qhat_{i} 
+ \beta\;\dfrac{\phat^2 \qhat_{i}+\qhat_{i}\,\phat^2}{2}
+ \beta'\,\dfrac{\phat_{i}\,(\phat\cdot\qhat) + (\qhat\cdot\phat)\,\phat_{i}}{2} 
\;,\cr
\phat_i & = & \phat_{i}\;.
\end{eqnarray}
The $\beta$ and $\beta'$ terms are symmetrized to ensure the hermiticity of $\xhat_i$.
Note that this representation allows us to write the Hamiltonian in terms of
canonical $\qhat_i$'s and $\phat_i$'s:
\begin{equation}
H'(\qbhat,\pbhat) \;\equiv\; H(\xbhat(\qbhat,\pbhat),\pbhat)\;.
\end{equation}
Thus, our deformation of the canonical commutation relations is mathematically
equivalent to a deformation of the Hamiltonian.\footnote{%
In this work, we do not address the question of whether the dependence of the Hamiltonian
on the position and momentum operators also need be modified in the presence of a minimal length.
Lacking in any guideline to do so, we simply keep them fixed to their standard forms.
}

By the standard replacements
\begin{equation}
\qhat_i\;=\;q_i\;,\quad \phat_i\;=\;\dfrac{\hbar}{i}\dfrac{\partial}{\partial q_i}\;,
\qquad\mbox{or}\qquad
\qhat_i\;=\;i\hbar\dfrac{\partial}{\partial p_i}\;,\quad \phat_i\;=\;p_i\;,
\end{equation}
$\xhat_i$ and $\phat_j$ can be represented as differential operators acting on a Hilbert
space of $L^2$ functions in either the $q_i$'s or the $p_i$'s, and 
one can write down a Schr\"odinger equation for a given Hamiltonian in either $q$-space or
$p$-space to solve for the energy eigenvalues.  
Note, however, that while the $p_i$'s are the eigenvalues of the momentum operators $\phat_i$,
 the $q_i$'s are not the eigenvalues of the position operator $\xhat_i$.
In fact, the existence of the minimal length implies that $\xhat_i$ cannot have any eigenfunctions
within either Hilbert spaces.  Therefore, the meaning of the wave-function in $q$-space is somewhat
ambiguous.  Nevertheless, the $q$-space representation is particularly useful when 
the Schr\"odinger equation cannot be solved exactly, since one can treat
\begin{equation}
\Delta H(\qbhat,\pbhat) \;=\; 
H'(\qbhat,\pbhat)
- H(\qbhat,\pbhat)
\end{equation}
as a perturbation and calculate the shifts in the energies via perturbation theory 
in $q$-space.

In the following, we look at the energy shifts induced by non-zero $\beta$ and $\beta'$
in the harmonic oscillator \cite{Kempf:1996fz,Chang:2001kn}, 
the Hydrogen atom \cite{Brau:1999uv,Benczik:2005bh}, 
and a particle in a uniform gravitational well \cite{Benczik:2007we,Brau:2006ca}.  
Since detailed derivations can be found in the respective references, 
we only provide an outline of the results in each case.

\subsubsection{Harmonic Oscillator}

\noindent
Consider a $D$-dimensional isotropic harmonic oscillator.
The Hamiltonian is of course
\begin{equation}
\Hhat\;=\; \dfrac{\hat{\pb}^2}{2m} + \dfrac{1}{2}m\omega^2\hat{\mathbf{x}}^2\;.
\label{Harmonic}
\end{equation}
The $p$-space representation of the operators are
\begin{eqnarray}
\xhat_i & = & i\hbar
\left[ (1 + \beta p^2)\frac{\partial}{\partial p_i}
       + \beta' p_i p_j \frac{\partial}{\partial p_j}
       + \left\{\beta+\beta'\left(\dfrac{D+1}{2}\right)-\delta(\beta+\beta')\right\} \,p_i
\right] \;,\cr
\phat_i & = & p_i \;.
\label{RepND}
\end{eqnarray}
Here, $\delta$ is an arbitrary real parameter which can be used to simplify the
representation of the operator $\xhat_i$ at the expense of modifying the definition 
of the inner product in $p$-space to
\begin{equation}
\langle f|g\rangle_\delta
\;=\; \int \dfrac{d^D\pb}{[\,1+(\beta+\beta')\pb^2\,]^\delta}\;f^*(\pb)\,g(\pb)\;.
\end{equation}
The introduction of $\delta$ is a canonical transformation which does not affect the energy eigenvalues \cite{Benczik:2007we}.  
The choice
\begin{equation}
\delta \;=\; \dfrac{\beta+\beta'\left(\dfrac{D+1}{2}\right)}{\beta+\beta'}
\end{equation}
eliminates the third term in the expression for $\xhat_i$.

\begin{figure}[ht]
\includegraphics[width=8cm]{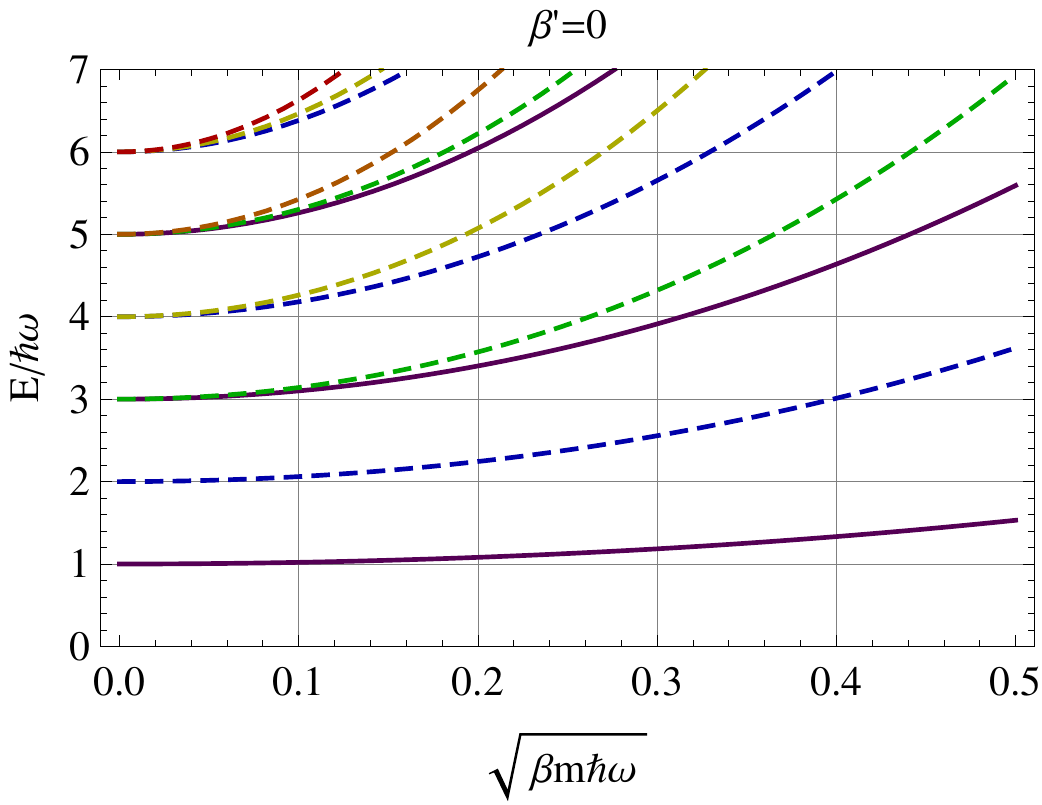}
\hspace{0.5cm}
\includegraphics[width=8cm]{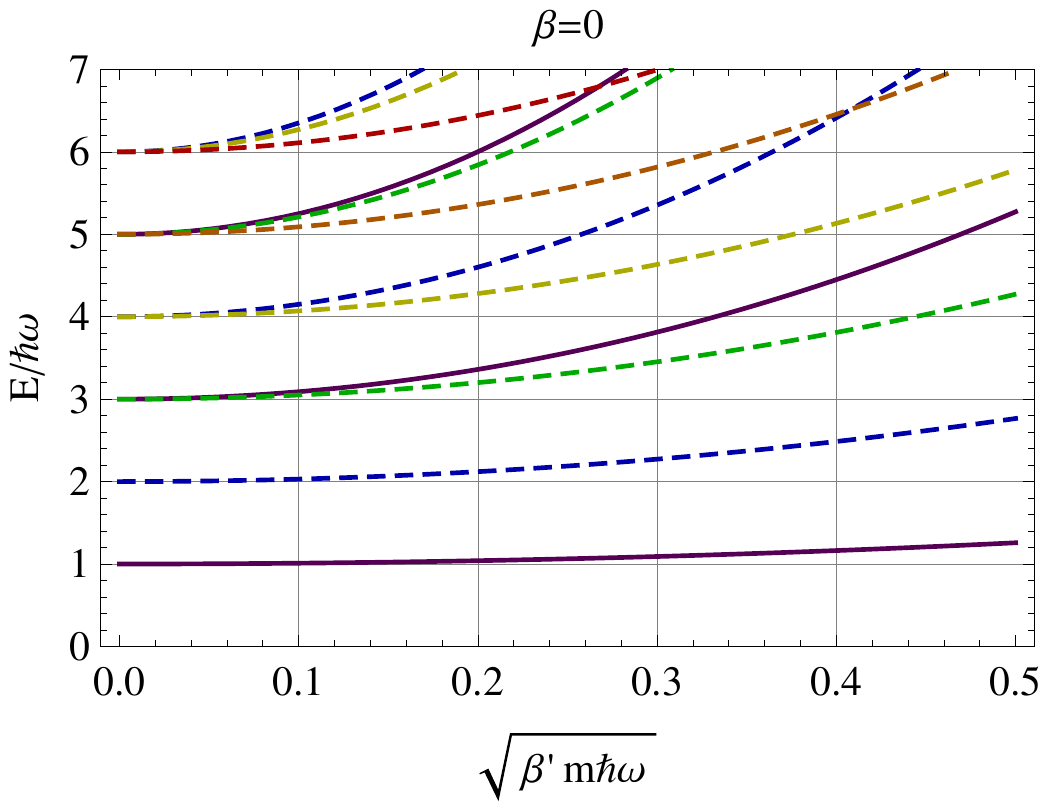}
\caption{The energy levels of the 2D isotropic harmonic oscillator for the cases
$\beta'=0$ (left) and $\beta=0$ (right). 
The purple solid lines indicate $s$-wave states which are singlets.
The dashed lines are doublets with the color indicating $\ell=1$ (blue),
$\ell=2$ (green), $\ell=3$ (yellow), $\ell=4$ (orange), and $\ell=5$ (red).
$\sqrt{\beta m\hbar\omega}$ is the ratio of the minimal length $\hbar\sqrt{\beta}$
to the characteristic length scale $\sqrt{\hbar/m\omega}$ of the system.
}
\label{HoEnergyLevels}
\end{figure}

The rotational symmetry of the Hamiltonian, Eq.~(\ref{Harmonic}), allows us to write
the wave-function in $p$-space as a product of 
a radial wave-function and a $D$-dimensional spherical harmonic:
\begin{equation}
\Psi_D(\pb) \;=\; R(p)\,Y_{\ell\,m_{D-2} m_{D-3}\cdots m_2 m_1}(\Omega)\;,\qquad
p\;\equiv\;|\pb|\;.
\end{equation}
The radial Schr\"odinger equation is then
\begin{eqnarray}
& & 
-m\hbar\omega
   \Biggl[
      \left\{ \Bigl[1 + (\beta + \beta') p^2\Bigr]\frac{\partial}{\partial p}
      \right\}^2
     +\frac{(D-1)(1+\beta\,p^2)[1 + (\beta + \beta') p^2]}{p}
      \frac{\partial}{\partial p}
\cr & & \hspace{5cm}
     -\frac{L^2(1+\beta\,p^2)^2}{p^2}
   \Biggr]\,R(p) 
+ \frac{1}{m\hbar\omega}\,p^2\,R(p)
\;=\; \frac{2E}{\hbar\omega}\,R(p)\;,
\label{Harmonic-Req}
\end{eqnarray}
where
\begin{equation}
L^2 \;=\; \ell(\ell+D-2)\;,\qquad \ell\;=\;0,1,2,\cdots
\end{equation}
is the eigenvalue of the angular momentum operator in $D$-dimensions.
The solution to Eq.~(\ref{Harmonic-Req}) has been worked out in detail in Ref.~\cite{Chang:2001kn},
and the energy eigenvalues are
\begin{eqnarray}
E_{n\ell}
& = & \hbar\omega
\left[
  \left( n + \frac{D}{2} \right) 
        \sqrt{ 1 + \left\{ \beta^2 L^2 
                         + \frac{ (D\beta + \beta')^2 }{ 4 }
                   \right\}m^2\hbar^2\omega^2
             }
\right. \cr
& & \qquad\left.
+ \left\{ (\beta + \beta')\left( n + \frac{D}{2}\right)^2
        + (\beta - \beta')\left(L^2 + \frac{D^2}{4}\right)
        + \beta'\frac{D}{2}
  \right\}\frac{m\hbar\omega}{2}
\right]\;,
\label{Enell}
\end{eqnarray}
with eigenfunctions given by
\begin{equation}
R_{n\ell}(p) \;=\; 
(\beta+\beta')^{D/4}
\sqrt{ \dfrac{ 2 (2k+a+b+1)\, k!\, \Gamma(k+a+b+1) }
             { \Gamma(k+a+1) \Gamma(k+b+1) }
     }\;
\left(\dfrac{1-z}{2}\right)^{\lambda/2}
\left(\dfrac{1+z}{2}\right)^{\ell/2}
P_{k}^{(a,b)}(z)\;.
\label{Rnell}
\end{equation}
Here, $P_{k}^{(a,b)}(z)$ is the Jacobi polynomial of order $k=(n-\ell)/2$
with argument
\begin{equation}
z 
  \;=\; \dfrac{ (\beta+\beta')p^2 -1 }{ (\beta+\beta')p^2 + 1 }\;,
\label{zdef}
\end{equation}
and
\begin{equation}
a \;=\; \dfrac{1}{m\hbar\omega(\beta+\beta')}
\sqrt{1+\left\{
\beta^2 L^2 + \dfrac{(D\beta+\beta')^2}{4}
\right\} m^2\hbar^2\omega^2}\;,\qquad
b \;=\; \frac{D}{2} + \ell - 1 \;,\qquad
\lambda \;=\;  \frac{D\beta+\beta'}{2(\beta+\beta')}+a
\;.
\end{equation}
Note that due to the $(n+D/2)^2$ dependend term in Eq.~(\ref{Enell}), the energy levels are no longer uniformly spaced.
Note also, that due to the explicit $L^2$-dependence, the original
\begin{equation}
\frac{ (D+n-1)! }{ (D-1)!\,n! } 
\end{equation}
fold degeneracy of the $n$-th energy level, which was due to states with different $k$ and $\ell$ sharing the
same $n=2k+\ell$, is resolved, leaving only the
\begin{equation}
  \frac{ (D+\ell-1)! }{ (D-1)!\,\ell! } 
- \frac{ (D+\ell-3)! }{ (D-1)!\,(\ell-2)! } 
\end{equation}
fold degeneracy for each value of $\ell$ due to rotational symmetry alone \cite{Chodos:1983zi}.
For example, in $D=2$ dimensions, the $(n+1)$-fold degeneracy of the $n$-th level
breaks down to the 2-fold degeneracies between the pairs of $m=\pm\ell$ states.  
This is illustrated in Fig.~\ref{HoEnergyLevels}

\subsubsection{Hydrogen Atom}

\begin{figure}[ht]
\includegraphics[width=8cm]{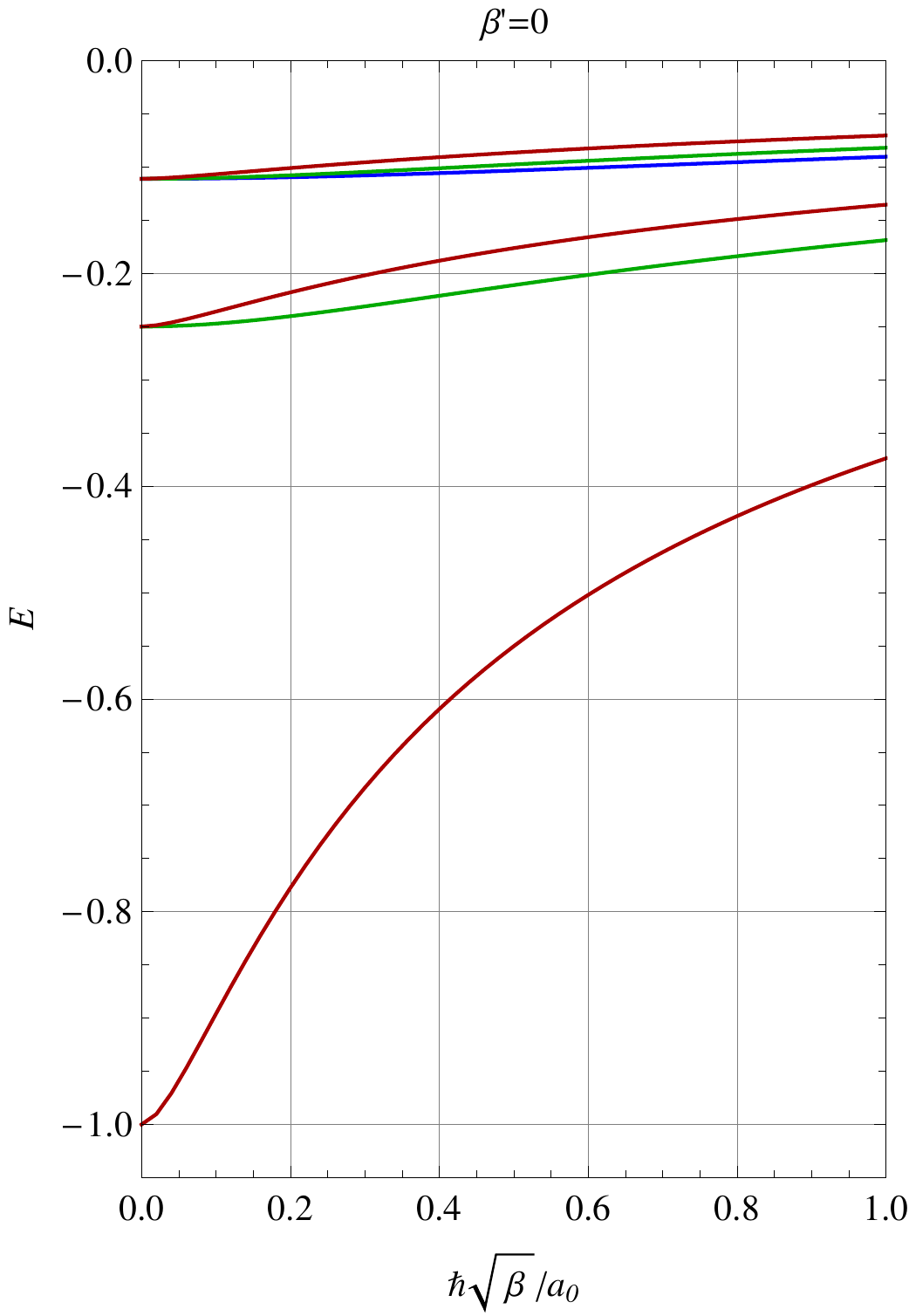}
\includegraphics[width=8cm]{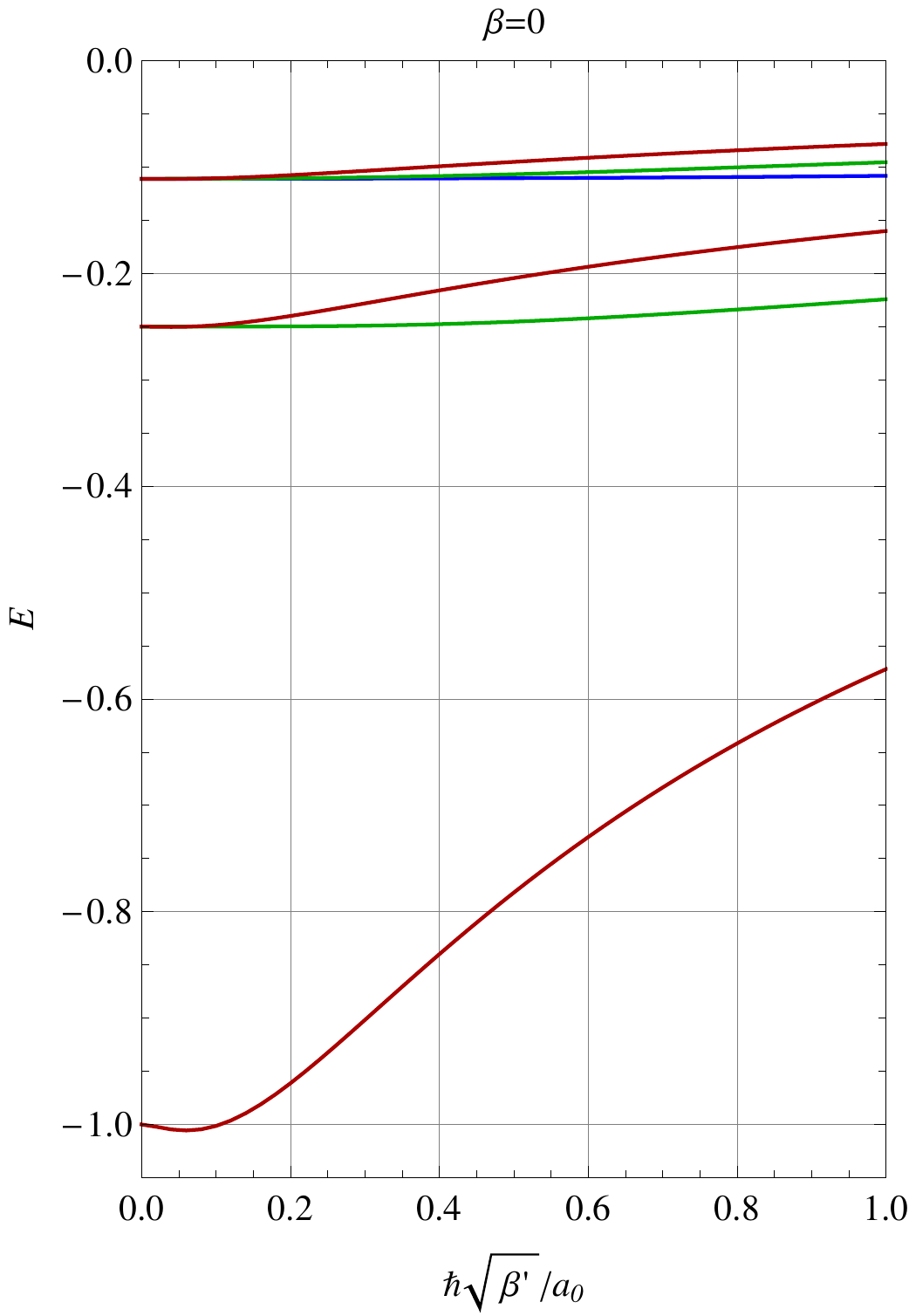}
\caption{Energy shifts of the $n=1$, $2$, and $3$ states of the Hydrogen atom for
the $\beta'=0$ (left) and $\beta=0$ (right) cases.
$a_0=\hbar^2/me^2$ is the Bohr radius and the energy is in units of the Rydberg constant $e^2/2a_0$.
The color of the lines indicate the orbital angular momentum: $s$ (red), $p$ (green), and $d$ (blue).
The $s$-wave states are affected non-perturbatively even for very small $\beta$ or $\beta'$, 
indicating their sensitivity to the singularity of the Coulomb potential at the origin.
}
\label{HatomEnergyLevels}
\end{figure}

The introduction of a minimal length to the Coulomb potential problem was first discussed by Born in 1933 \cite{born:1933}.
There, it was argued that the singularity at $r=0$ will be blurred out.  Here, we find a similar effect.
We consider the usual Hydrogen atom Hamiltonian in $D$-dimensions:
\begin{equation}
\Hhat\;=\;\dfrac{\hat{\pb}^2}{2m} - \dfrac{e^2}{\rhat}\;,
\end{equation}
where the operator $1/\rhat$ is defined as the inverse of the square-root of 
the operator
\begin{equation}
\rhat^2 \;=\; \sum_{i=1}^{D} \xhat_i^2\;. 
\end{equation}
$1/\rhat$ will be best represented in the basis in which $\rhat^2$ is diagonal.
The eigenvalues of $\rhat^2$ can be obtained from those of the
harmonic oscillator, Eq.~(\ref{Enell}), by taking the limit $m\rightarrow\infty$:
\begin{eqnarray}
r_{k\ell}^2 
& = & \lim_{m\rightarrow\infty} \dfrac{2E_{n\ell}}{m\omega^2} \cr
& = & \hbar^2(\beta+\beta')
\left[
\left\{\left(2k+\ell+\dfrac{D}{2}\right)+
\dfrac{1}{\beta+\beta'}\sqrt{\beta^2 L^2 + \dfrac{(D\beta+\beta')^2}{4}}
\right\}^2
-\dfrac{\beta'}{\beta+\beta'}
\Biggl\{ L^2 + \dfrac{(D-1)^2}{4} \Biggr\}
\right]\;.
\end{eqnarray}
The corresponding eigenfunctions are given by the same expression as Eq.~(\ref{Rnell})
except with $a$ replaced by
\begin{equation}
a \;=\; \dfrac{1}{\beta+\beta'}
\sqrt{
\beta^2 L^2 + \dfrac{(D\beta+\beta')^2}{4}
}
\;.
\end{equation}
Denoting these eigenfunctions as $R_{k\ell}(p)$, we can define
\begin{equation}
\dfrac{1}{\rhat}\,R_{k\ell}(p)\;=\;\dfrac{1}{r_{k\ell}}\,R_{k\ell}(p)\;.
\end{equation}
As in the harmonic oscillator case, the rotational symmetry of the Hamiltonian
allows us to write an energy eigenstate wave-function as a product of a radial wave-function and a spherical harmonic.
The radial wave-function can then be expressed as a superposition of the $\hat{r}^2$ eigenfunctions
with fixed $\ell$:
\begin{equation}
R_\ell(p) \;=\;\sum_{k=0}^\infty f_k R_{k\ell}(p)\;.
\end{equation}
The radial Schr\"odinger equation will impose a recursion relation on the coefficients
$f_n$, which can be solved numerically on a computer.
The condition that the resulting function be square-integrable determines the eigenvalues $E$.
The detailed procedure can be found in Ref.~\cite{Benczik:2005bh,Benczik:2007we}.
Here, we only display the results for the $D=3$ case in Fig.~\ref{HatomEnergyLevels}.
As can be seen, the degeneracy between difference angular momentum states are lifted,
just as in the harmonic oscillator case.

It is also possible to calculate the energy shifts perturbatively using the $q$-space
representation for the cases $D\ge 4$ or $\ell\neq 0$.
The unperturbed energy eigenfunctions in $D$-dimensions are
\begin{equation}
R_{n\ell}(q) \;=\; 
\sqrt{\dfrac{2^{2D}}{a_0^D\left(2n+D-3\right)^{D+1}}\,
\dfrac{(n-\ell-1)!}{(n+\ell+D-3)!}}
\;e^{-\rho/2} \rho^\ell L_{n-\ell-1}^{(2\ell+D-2)}(\rho)\;,
\end{equation}
where $a_0 = \hbar^2/me^2$ is the Bohr radius,
$L_k^{(\lambda)}(\rho)$ the order $k$ Laguerre polynomial, 
and
\begin{equation}
\rho\;=\;\dfrac{2q}{a_0\left(n+\frac{D-3}{2}\right)}
\;.
\end{equation}
The eigenvalues are
\begin{equation}
E_{n} \;=\; - \dfrac{e^2}{2a_0\left(n+\frac{D-3}{2}\right)^2}
\;,\qquad\qquad
n\,=\,1,\,2,\,3,\cdots
\end{equation}
The operator $1/\rhat$ can be expanded in powers of $\beta$ and $\beta'$ 
as \cite{Benczik:2007we}
\begin{eqnarray}
\dfrac{1}{\rhat}
\;=\; \dfrac{1}{q} 
+\hbar^2\beta
\left( \dfrac{1}{q}\dfrac{\partial^2}{\partial q^2} 
+ \dfrac{D-2}{q^2}\dfrac{\partial}{\partial q}
- \dfrac{L^2+D-2}{q^3}
\right)
+\hbar^2\beta'
\left( \dfrac{1}{q}\dfrac{\partial^2}{\partial q^2} 
+ \dfrac{D-2}{q^2}\dfrac{\partial}{\partial q}
+ \dfrac{D^2-5D+8}{4q^3}
\right)
+ \cdots
\;,
\label{rinvexpansion}
\end{eqnarray}
and the expectation value of the extra terms converges
for $\ell\neq 0$ or $D\ge 4$, yielding
\begin{equation}
\Delta E_{n\ell}
\;=\; \dfrac{e^2}{a_0\left(n+\frac{D-3}{2}\right){}^3}\;
\dfrac{\hbar^2}{a_0^2}\!
\left[
\dfrac{(D-1)(2\beta-\beta')}{4\left(\ell+\frac{D-3}{2}\right)\left(\ell+\frac{D-2}{2}\right)\left(\ell+\frac{D-1}{2}\right)}
+\dfrac{(2\beta+\beta')}{\left(\ell+\frac{D-2}{2}\right)}
-\dfrac{(\beta+\beta')}{\left(n+\frac{D-3}{2}\right)}
\right]
\;,
\end{equation}
which agrees very well with the numerical results for all cases to which it is applicable.
For $D=3$, this formula reduces to
\begin{equation}
\Delta E_{n\ell}
\;=\; \dfrac{e^2}{a_0 n^3}\;
\dfrac{\hbar^2}{a_0^2}\!
\left[
\dfrac{(2\beta-\beta')}{2\,\ell\left(\ell+\frac{1}{2}\right)\left(\ell+1\right)}
+\dfrac{(2\beta+\beta')}{\left(\ell+\frac{1}{2}\right)}
-\dfrac{(\beta+\beta')}{n}
\right]
\;,
\end{equation} 
which is clearly problematic for $\ell=0$.
This is due to the breakdown of the expansion Eq.~(\ref{rinvexpansion}) near $q=0$ for $D\le 3$.
Physically, this can be interpreted to mean that the $s$-wave in 3D and lower dimensions is sensitive to the 
non-perturbative resolution of the singularity at the origin due to the minimal length.  
Interestingly, in 4D and higher, there are enough spatial dimensions for the wave-function to spread out
around the origin so that even the $s$-wave is insensitive to the singularity, and
the effect of the minimal length becomes perturbative.

\subsubsection{Uniform Gravitational Potential}

\begin{figure}[ht]
\includegraphics[width=8cm]{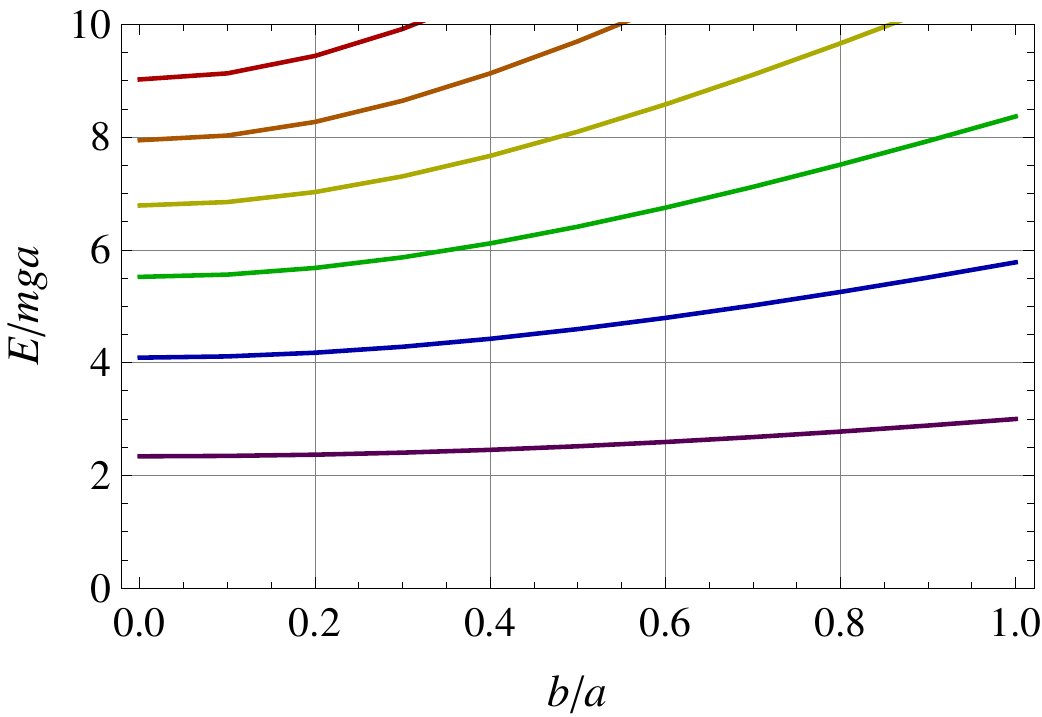}
\caption{The $b$-dependence of the lowest energy levels of a particle of mass $m$ in 
a linear gravitaional potential $V(x)=mgx$ with $x>0$.
$a=[\hbar^2/2m^2 g]^{1/3}$ is the characteristic length scale of the system,
and $b=\hbar\sqrt{\beta}$ is the minimal length.}
\label{mgzEnergyLevels}
\end{figure}

This subsection is based on unpublished material by Benczik in Ref.~\cite{Benczik:2007we}.
Consider the 1D motion of a particle in a linear potential 
\begin{equation}
V(x) \;=\; \left\{\begin{array}{ll}
mgx\qquad & x\;>\;0\;,\\
\infty    & x\;\le\;0\;.
\end{array}\right.
\end{equation}
The Hamiltonian is
\begin{equation}
\Hhat \;=\; \dfrac{\phat^2}{2m} + mg\,\xhat\;.
\end{equation}
Since $\xhat$ does not have any eigenstates within the Hilbert space,
the condition $x>0$ is replaced by $\vev{\xhat}> 0$.
In the $q$-space representation, the operators are given by
\begin{eqnarray}
\xhat & = & q\left(1-\hbar^2\beta\dfrac{d^2}{dq^2}\right)\;,\cr
\phat & = & \dfrac{\hbar}{i}\,\dfrac{d}{d q}\;,
\end{eqnarray}
and the Schr\"odinger equation becomes
\begin{equation}
\hat{H}\psi \;=\; -\dfrac{\hbar^2}{2m}\,\dfrac{d^2\psi}{dq^2}
+ mg\,q\left(1-\hbar^2\beta\dfrac{d^2}{dq^2}\right)\psi
\;=\; E\,\psi\;.
\label{mgxEq}
\end{equation}
The condition $\vev{\xhat}> 0$ can be imposed by restricting the domain of $q$ to $q>0$, and
demanding that the wave function vanish at $q=0$. 
The solution to the $\beta=0$ case is given by the Airy function
\begin{equation}
\psi_n(q) \;=\; \dfrac{1}{|\Ai'(\alpha_n)|}\;\Ai\left(\dfrac{q}{a} + \alpha_n\right)\;,\qquad\qquad
a\,=\,\left[\dfrac{\hbar^2}{2m^2 g}\right]^{1/3}\;,
\end{equation}
with eigenvalues
\begin{equation}
\dfrac{E_n}{mga} \;=\; -\alpha_n\;,
\end{equation}
where
\begin{equation}
\cdots \;<\;\alpha_3\;<\;\alpha_2\;<\;\alpha_1\;<\;0
\end{equation}
are the zeroes of $\Ai(z)$. 
The solution to the $\beta\neq 0$ case is given in terms of the
confluent hypergeometric function of the second kind \cite{Ufunction}:
\begin{equation}
\psi(q) \;\propto\; e^{-q/b}\;
U\left(-\dfrac{1}{2}\!\left[\dfrac{E}{mgb}+\dfrac{a^3}{b^3}\right];\,0\,;\,
2\!\left[\dfrac{a^3}{b^3}+\dfrac{q}{b}\right]\,\right)\;,
\qquad
a\,=\,\left[\dfrac{\hbar^2}{2m^2 g}\right]^{1/3}\;,\quad
b\,=\,\hbar\sqrt{\beta}\;.
\end{equation}
The energy eigenvalues are determined by the condition
\begin{equation}
U\left(-\dfrac{1}{2}\!\left[\dfrac{E}{mgb}+\dfrac{a^3}{b^3}\right];\,0\,;\,\dfrac{2a^3}{b^3}\,\right)\;=\;0\;,
\end{equation}
which can be solved numerically using Mathematica.
In Fig.~\ref{mgzEnergyLevels}, we plot the $b$-dependence of the energies of the lowest lying states.
The energies of higher-dimensional cases, in which there are one or more spatial dimensions orthogonal to the
potential direction, are discussed in Ref.~\cite{Benczik:2007we,Brau:2006ca}.

\subsection{Experimental Constraints}

As these three examples show, the main effect of the introduction of the 
minimal length into quantum mechanical systems is the shifts in energy levels
which also leads to the breaking of well known degeneracies.
The natural question arises whether these shifts can be used to
constrain the minimal length experimentally.  
Of course, if the minimal length is at the Planck scale,
detecting its actual effect would be impossible.
However, the exercise is of interest to 
models of large extra dimensions which possess a much lower
effective Planck scale than the 4D value \cite{LargeExtraDimensions}.

In the case of the harmonic oscillator, actual physical systems
are never completely harmonic, so it is difficult to distinguish
the shift in energy due to anharmonicity with that due to a possible
minimal length.  Ref.~\cite{Chang:2001kn} considers using the
energy levels of an electron in a Penning trap to constrain
$\beta$, and finds that even under highly optimistic and unrealistic
assumptions, the best bound that can be hoped for is
\begin{equation}
\dfrac{1}{\sqrt{\beta}}\;\agt\;1\;\mathrm{GeV}/c\;.
\label{hobetabound}
\end{equation}

Refs.~\cite{Benczik:2007we,Benczik:2005bh} consider placing a bound on 
$\beta$ using the $1S$ Lamb shift of the Hydrogen atom.
The current best experimental value is that given by Schwob et al. in \cite{Schwob:1999zz}:
\begin{equation}
L_{1s}^\mathrm{exp} 
\;=\; 8172.837(22)\;\mathrm{MHz}\;.
\end{equation}
This is to be compared with the theoretical value, for which we use that
given in Ref.~\cite{Mallampalli:1998zza}:
\begin{equation}
L_{1s}^\mathrm{th}
\;=\; 8172.731(40)\;\mathrm{MHz}\;.
\end{equation}
The calculation requires the experimentally determined proton rms charge radius $r_p$
as an input, and the error on $L_{1s}^\mathrm{th}$ is dominated by the experimental error on $r_p$.
Here, the value of $r_p = 0.862(12)\;\mathrm{fm}$ \cite{Simon:1980hu} was used.
Attributing the entire discrepancy to $\beta$ ($\beta'=0$), 
Refs.~\cite{Benczik:2007we,Benczik:2005bh} cite:
\begin{equation}
\dfrac{1}{\sqrt{\beta}}\;\agt\; 7\,\mathrm{GeV}/c\;,
\end{equation}
which is only slightly better than Eq.~(\ref{hobetabound}).
The is no bound on $\beta'$ ($\beta=0$) since the shift is in the wrong direction as can be seen in
Fig.~(\ref{HatomEnergyLevels}).

The energy levels of neutrons in a linear gravitational potential have been
measured by Nesvizhevsky et al. \cite{Nesvizhevsky:2002ef}.
However, as analyzed by Brau and Buisseret \cite{Brau:2006ca},
the experimental precision is still very many orders of magnitude above what is necessary
to place a meaningful bound on $\beta$. 
The current lower bound on $1/\sqrt{\beta}$ is on the order of $100\,\mathrm{eV}/c$.

\section{Classical Limit -- The Liouville Theorem and the Density of States}

Note that rewriting our 1D deformed commutator as
\begin{equation}
[\,\xhat,\,\phat\,]\;=\;i\hbar A(\phat^2)
\end{equation}
suggests that $\hbar A(p^2)$ takes on the role of a momentum dependent Planck constant.
Given that $\hbar$ determines the size of a quantum mechanical state in phase space,
a momentum dependent $\hbar$ would imply that the size of this state must 
scale according to $A(p^2)$ as it evolves.
To see whether this interpretation makes sense, we formally take the naive classical limit by 
replacing commutators with Poisson brackets,
\begin{equation}
\dfrac{1}{i\hbar}[\,\xhat,\,\phat\,]\;=\;A(\phat^2)
\qquad\longrightarrow\qquad
\{\,x,\,p\,\}\;=\;A(p^2)\;,
\end{equation}
and proceed to derive the analogue of Liouville's theorem \cite{Chang:2001bm}.
The Poisson brackets among the $x_i$'s and $p_i$'s for the multidimensional case are
\begin{eqnarray}
\{\,x_i,\,p_j\,\} & = & A\,\delta_{ij}+B\,p_i p_j\;,\cr
\{\,p_i,\,p_j\,\} & = & 0\;,\cr
\{\,x_i,\,x_j\,\} & = & -\left[\dfrac{2(A+B\,\pb^2)}{A}\,\dfrac{dA\;}{d\pb^2}-B\right] \left(x_i p_j - x_j p_i\right)\;.
\end{eqnarray}
The generic Poisson bracket of arbitrary functions of the coordinates and momenta can then be defined as
\begin{equation}
\{F,G\} = 
\left( \frac{\partial F}{\partial x_i}
       \frac{\partial G}{\partial p_j}
     - \frac{\partial F}{\partial p_i}
       \frac{\partial G}{\partial x_j}
\right) \{ x_i, p_j \}
+ \frac{\partial F}{\partial x_i}
  \frac{\partial G}{\partial x_j}
  \{ x_i, x_j \}\;.
\label{Eq:Poisson2}
\end{equation}
Here, we use the convention that repeated indices are summed.
Assuming that the equations of motion of $x_i$ and $p_i$ are given formally by:
\begin{eqnarray}
\dot{x}_i & = & \{\,x_i,\,H\,\}\;=\;\phantom{-}\{\,x_i,\,p_j\,\}\dfrac{\partial H}{\partial p_j}
+ \{\,x_i,\,x_j\,\}\dfrac{\partial H}{\partial x_j}
\;,\cr
\dot{p}_i & = & \{\,p_i,\,H\,\}\;=\;-\{\,x_j,\,p_i\,\}\dfrac{\partial H}{\partial x_j}
\;,
\end{eqnarray}
the evolution of $x_i$ and $p_i$ during an infinitesimal time interval $\delta t$ is found to be:
\begin{eqnarray}
x'_i & = & x_i + \dot{x}_i\,\delta t \;=\; x_i 
+\left[ \{\,x_i,\,p_j\,\}\dfrac{\partial H}{\partial p_j}
+ \{\,x_i,\,x_j\,\}\dfrac{\partial H}{\partial x_j}
\right]\delta t\;,\cr
p'_i & = & p_i + \dot{p}_i\,\delta t \;=\; p_i - \{\,x_j,\,p_i\,\}\dfrac{\partial H}{\partial x_j}\delta t\;.
\end{eqnarray}
To find the change in phase space volume associated with this evolution, we calculate the Jacobian 
of the transformation from $(x_1,x_2,\cdots,x_D;p_1,p_2,\cdots,p_D)$ to
$(x'_1,x'_2,\cdots,x'_D;p'_1,p'_2,\cdots,p'_D)$:  
\begin{equation}
d^D\xb'\,d^D\pb' \;=\;
\left|
\dfrac{\partial(x'_1,x'_2,\cdots,x'_D;p'_1,p'_2,\cdots,p'_D)}{\partial(x_1,x_2,\cdots,x_D;p_1,p_2,\cdots,p_D)}
\right|
d^D\xb\,d^D\pb \;.
\end{equation}
Since
\begin{equation}
\begin{array}{ll}
\dfrac{\partial x'_i}{\partial x_j} \;=\; 
\delta_{ij} + \dfrac{\partial\dot{x}_i}{\partial x_j}\,\delta t\;,
\qquad & 
\dfrac{\partial x'_i}{\partial p_j} \;=\;
\dfrac{\partial\dot{x}_i}{\partial p_j}\,\delta t\;,
\\
\dfrac{\partial p'_i}{\partial x_j} \;=\;
\dfrac{\partial\dot{p}_i}{\partial x_j}\,\delta t\;,
\qquad &
\dfrac{\partial p'_i}{\partial p_j} \;=\; 
\delta_{ij} + \dfrac{\partial\dot{p}_i}{\partial p_j}\,\delta t\;,
\end{array}
\end{equation}
we find:
\begin{equation}
\left|
\dfrac{\partial(x'_1,x'_2,\cdots,x'_D;p'_1,p'_2,\cdots,p'_D)}{\partial(x_1,x_2,\cdots,x_D;p_1,p_2,\cdots,p_D)}
\right|
\;=\; 1 + \left(\dfrac{\partial\dot{x}_i}{\partial x_i}+\dfrac{\partial\dot{p}_i}{\partial p_i}\right)\delta t
+ O(\delta t^2)\;,
\end{equation}
where
\begin{eqnarray}
\dfrac{\partial\dot{x}_i}{\partial x_i} + \dfrac{\partial\dot{p}_i}{\partial p_i}
& = & \dfrac{\partial}{\partial x_i}\!
\left[ \{\,x_i,\,p_j\,\}\dfrac{\partial H}{\partial p_j}
+ \{\,x_i,\,x_j\,\}\dfrac{\partial H}{\partial x_j}
\right] 
+
\dfrac{\partial}{\partial p_i}\!
\left[ -\{\,x_j,\,p_i\,\}\dfrac{\partial H}{\partial x_j} \right]
\cr
& = & 
  \dfrac{\partial}{\partial x_i}\!\Bigl[ \{\,x_i,\,x_j\,\} \Bigr] \dfrac{\partial H}{\partial x_j}
- \dfrac{\partial}{\partial p_i}\!\Bigl[\{\,x_j,\,p_i\,\}\Bigr] \dfrac{\partial H}{\partial x_j}
\cr
& = & -(D-1)\left[\dfrac{2(A+B\,\pb^2)}{A}\,\dfrac{dA\;}{d\pb^2}-B\right]p_j\dfrac{\partial H}{\partial x_j}
-\left[2\,\dfrac{dA\;}{d\pb^2}+2\,\dfrac{dB\;}{d\pb^2}\pb^2 + (D+1)B\right]p_j\dfrac{\partial H}{\partial x_j}
\cr
& = & -\left[(D-1)\left(\dfrac{2(A+B\,\pb^2)}{A}\dfrac{dA\;}{d\pb^2}\right)
+2\left(\dfrac{dA\;}{d\pb^2}
+\dfrac{dB\;}{d\pb^2}\pb^2 + B
\right)
\right]p_j\dfrac{\partial H}{\partial x_j}\;.
\label{Jacobian}
\end{eqnarray}
On the other hand, using
\begin{equation}
\delta\pb^2 
\;=\; 2p_i \delta p_i
\;=\; 2p_i\,\dot{p}_i\,\delta t
\;=\; - 2\,(A+B\pb^2)\; p_j \dfrac{\partial H}{\partial x_j}\,\delta t\;,
\end{equation}
we have
\begin{eqnarray}
A' & = & 
A + \dfrac{dA\;}{d\pb^2}\delta\pb^2
\cr
& = & A
\left[ 1 - 
\left( \dfrac{2(A+B\pb^2)}{A}\,\dfrac{dA\;}{d\pb^2}
\right) p_j \dfrac{\partial H}{\partial x_j}\,\delta t
\right]
\;,\cr
A'+B'\pb^{\prime 2} & = & 
(A + B\pb^2) +
\left( \dfrac{dA\;}{d\pb^2} + \dfrac{dB\;}{d\pb^2}\,\pb^2 + B\right)\delta\pb^2 
\cr
& = &
(A+B\pb^2)
\left[ 1  - 
2\left( \dfrac{dA\;}{d\pb^2} + \dfrac{dB\;}{d\pb^2}\,\pb^2 + B\right)
p_j \dfrac{\partial H}{\partial x_j}\,\delta t \right]
\;,
\end{eqnarray}
where we have used the shorthand $A'=A(\pb^{\prime 2})$ and $B'=B(\pb^{\prime 2})$.
Thus
\begin{equation}
\dfrac{(A')^{D-1}(A'+B'\pb^{\prime 2})}
      {A^{D-1}(A+B\pb^2)}
\;=\;
\Biggl[\;1 -\left\{(D-1)\left(\dfrac{2(A+B\,\pb^2)}{A}\dfrac{dA\;}{d\pb^2}\right)
+2\left(\dfrac{dA\;}{d\pb^2}
+\dfrac{dB\;}{d\pb^2}\pb^2 + B
\right)
\right\} p_j\dfrac{\partial H}{\partial x_j}\delta t
\;\Biggr]
\;.
\label{AAB}
\end{equation}
Comparing Eqs.~(\ref{Jacobian}) and (\ref{AAB}), it is clear that
the ratio
\begin{equation}
\dfrac{d^D\mathbf{x}\,d^D\pb}{A^{D-1}(A+B\pb^2)}
\label{InvariantPSVolume}
\end{equation}
is invariant under time evolution.

This behavior of the phase space volume can be demonstrated using simple Hamiltonians.
In Ref.~\cite{Benczik:2002tt}, we solve the harmonic oscillator, and coulomb potential problems
for the case $A=1+\beta\pb^2$ and $B=\beta'$. 
There, in addition to the behavior of the phase space, it is found that the orbits
of particles in these potentials no longer close on themselves.
This is consistent with the breaking of degeneracies observed in the quantum cases
which are associated with the conservation of the Runge-Lenz vector. 

For the case $B=0$, Eq.~(\ref{InvariantPSVolume}) reduces to $d^D\xb\,d^D\pb/A^D$,
and our intepretation of $\hbar A(p^2)$ as the momentum dependent Planck constant
which determines the size of a unit quantum cell becomes apparent.
Integrating Eq.~(\ref{InvariantPSVolume}) over space, 
\begin{equation}
\dfrac{1}{V}\int \dfrac{d^D\xb\,d^D\pb}{A^{D-1}(A+B\pb^2)}
\;=\; \dfrac{d^D\pb}{A^{D-1}(A+B\pb^2)}
\;,
\end{equation}
we can identify
\begin{equation}
\rho(\pb^2)
\;=\;\dfrac{1}{A^{D-1}(A+B\pb^2)}
\label{DensityOfStates}
\end{equation}
as the density of states in momentum space.
At high momentum where $A$ and $B\pb^2$ become large, $\rho(\pb^2)$ will be suppressed.
We look at the impact of this suppression on the cosmological constant problem next.

\section{Vacuum energy and the minimal length}

\subsection{The Cosmological Constant and the Density of States}

The origin of the cosmological constant $\Lambda = 3H_0^2\Omega_\Lambda$ remains a mystery, 
and its understanding presents a major challenge to theoretical physics \cite{CosmoConstant}.
It is a contentious issue for string theory, 
notwithstanding its being the leading candidate for quantum gravity,
though various hints exist that may point towards its resolution \cite{Banks:2004zb,Polchinski:2006gy}.
Furthermore, the problem has recently assumed added urgency
due to observations that the cosmological constant is small, positive, and clearly non-zero \cite{Copeland:2006wr}.
In terms of the parameter $\Omega_\Lambda$, the most up to date value is $\Omega_\Lambda \sim 0.73$. 
With the Hubble parameter $h\sim 0.7$,\footnote{%
The parameter $h$ is defined as $h=H_0/(100\,\mathrm{km/s/Mpc})$.}
we obtain as the vacuum energy density
\begin{eqnarray}
\dfrac{c^2\Lambda}{8\pi\GN} 
& = & c^2\rho_\mathrm{crit}\,\Omega_\Lambda 
\;=\; \left(\dfrac{3H_0^2 c^2}{8\pi\GN}\right)\Omega_\Lambda 
\;=\; (8.096\times 10^{-47}\,\mathrm{GeV}^4/\hbar^3 c^3)(\Omega_\Lambda h^2) 
\;\sim\; 10^{-47}\,\mathrm{GeV}^4/\hbar^3 c^3
\;. 
\end{eqnarray}
The order of magnitude of this result is set by the dimensionful prefactor
in the parentheses
which can be expressed in terms of the Planck length 
$\ellP = \hbar/\muP = \sqrt{\hbar\GN/c^3} \sim 10^{-35}\,\mathrm{m}$, 
and the scale of the visible universe 
$\ell_0 = \hbar/\mu_0 \equiv c/H_0 \sim 10^{26}\,\mathrm{m}$ as
\begin{equation}
\dfrac{H_0^2 c^2}{\GN}
\;=\; \dfrac{c}{\hbar^3}\,\muP^2\mu_0^2
\;=\; \dfrac{\hbar c}{\ellP^2\,\ell_0^2}
\;.
\end{equation}

In quantum field theory (QFT),
the cosmological constant is calculated as the sum of the vacuum fluctuation energies of
all momentum states. This is clearly infinite, so the integral is usually cut off at the Planck scale 
$\muP = \hbar/\ellP$ beyond which spacetime itself is expected to become foamy \cite{Wheeler:1957mu},
and the calculation untrustworthy.  
For a massless particle, we find:
\begin{equation}
\dfrac{1}{(2\pi\hbar)^3}
\int^{\muP} d^3\pb\left[\dfrac{1}{2}\hbar\omega_p\right]
\;=\; \dfrac{c}{4\pi^2\hbar^3}\int_0^{\muP} dp\;p^3
\;=\; \dfrac{c}{16\pi^2\hbar^3}\,\muP^4
\;=\; \dfrac{\hbar c}{16\pi^2}\;\dfrac{1}{\ellP^4}
\;\sim\; 10^{74}\,\mathrm{GeV}^4/\hbar^3 c^3
\;,
\label{LambdaUsualCase}
\end{equation}
which is about 120 orders of magnitude above the measured value.  
Note that this difference is essentially a factor of $(\ell_0/\ellP)^2$,
the scale of the visible universe in Planck units squared.
The change in the density of states suggested by the MLUR would change this calculation to
\begin{equation}
\dfrac{1}{(2\pi\hbar)^3}
\int^\infty d^3\pb\;\rho(\pb^2)\!\left[\dfrac{1}{2}\hbar\omega_p\right]
\;=\; \dfrac{c}{4\pi^2\hbar^3}
\int_0^\infty dp\;\dfrac{p^3}{A(p^2)^2[A(p^2)+p^2 B(p^2)]}\;.
\end{equation}
For the case $A(p^2)=1+\beta p^2$, $B(p^2)=0$, we find \cite{Chang:2001bm}:
\begin{equation}
\dfrac{c}{4\pi^2\hbar^3}\int_0^\infty dp\;\dfrac{p^3}{(1+\beta p^2)^3}
\;=\; \dfrac{c}{16\pi^2\hbar^3\beta^2}
\;=\; \dfrac{c}{16\pi^2\hbar^3}\,\mu_s^4
\;=\; \dfrac{\hbar c}{16\pi^2}\;\dfrac{1}{\ell_s^4}
\;,\qquad\quad
\ell_s \,=\, \dfrac{\hbar}{\mu_s} \,=\, \hbar\sqrt{\beta}\;.
\label{LambdaMLUR}
\end{equation}
The integral is finite, without a UV cutoff, due to the suppression of the contribution of high momentum states.\footnote{%
There is an intriguing similarity here with Planck's resolution of the UV catastrophe of the black body radiation.}
However, if we make the identification $\ell_s=\ellP$, then this result is identical to
Eq.~(\ref{LambdaUsualCase}) and nothing is gained.  
Of course, this is not surprising given that $\ell_s$ is the only scale in the calculation,
and effectively plays the role of the UV cutoff.
To obtain the correct value of the cosmological constant from the above expression, we must choose
$\ell_s \sim \sqrt{\ellP\ell_0} \sim 10^{-5}\,\mathrm{m}$, which is too large to be the minimal length,
or equivalently, $\mu_s = \hbar/\ell_s \sim \sqrt{\muP\mu_0} \sim 10^{-3}\,\mathrm{eV}/c$,
which is too small to be the UV cutoff.
However, we mention in passing that $\sqrt{\ellP\ell_0}$ can be considered 
the uncertainty in measuring $\ell_0$ due to the foaminess of spacetime \cite{Wheeler:1957mu,Wigner:1957ep}, and 
has been argued as the possible size of a spacetime quantum cell when quantum gravity is
properly taken into account \cite{ng,AmelinoCamelia:1994vs,Diosi:1989hy}. 
At the moment, this point of view seems difficult to reconcile with phenomenological considerations.

We could introduce a second scale into the problem by letting
$B(p^2)=\beta'\neq 0$.  This leads to
\begin{eqnarray}
\dfrac{c}{4\pi^2\hbar^3}\int_0^\infty dp\;\dfrac{p^3}{(1+\beta p^2)^2[1+(\beta+\beta')p^2]}
& = & \dfrac{c}{8\pi^2\hbar^3}\;\dfrac{1}{\beta\beta'}
\left[ 1- \dfrac{\beta}{\beta'}\ln\left(1+\dfrac{\beta'}{\beta}\right)
\right] \cr
& \xrightarrow{\beta'\gg\beta} &
\dfrac{c}{8\pi^2\hbar^3}\;\dfrac{1}{\beta\beta'}
\;=\; \dfrac{c}{8\pi^2\hbar^3}\;\mu_s^2 \mu_s^{\prime 2}
\;=\; \dfrac{\hbar c}{8\pi^2}\;\dfrac{1}{\ell_s^2\ell_s^{\prime 2}}
\;,
\label{LambdaMLUR2}
\end{eqnarray}
where $\ell'_s = \hbar/\mu'_s = \hbar\sqrt{\beta'}$.
If we identify $\ell_s=\ellP$, then we must have $\ell'_s\sim\ell_0$, which is even more problematic
than $\sqrt{\ellP\ell_0}$.

As these considerations show, our simple choice for $A(p^2)$ and $B(p^2)$ succeeds in
rendering the cosmological constant finite, but does not provide an adequate suppression.
Would some other choice of $A(p^2)$ and $B(p^2)$ do better?
To this end, let us try to see whether we can reverse engineer these functions so that the correct
order of magnitude is obtained.  Let us write
\begin{equation}
\epsilon^4 \;=\; \int_0^\infty dp\;\rho(p^2)\,p^3\;. 
\end{equation}
To generate the correct value for the cosmological constant, we must have 
$\epsilon \sim \sqrt{\muP\mu_0} = 10^{-3}\mathrm{eV}/c$, as we have seen.
At this point, we invoke some numerology and note that if
the SUSY breaking scale $\muSUSY$ is on the order of a few $\mbox{TeV}/c$, then
the seesaw formula,
\begin{equation}
\epsilon \;\sim\; \frac{\muSUSY^2}{\muP}  \;\sim\; 10^{-3}\,\mathrm{eV}/c\;,
\label{seesaw}
\end{equation}
would give the correct size for $\epsilon$ as observed by Banks \cite{Banks:2001zj}.
This expression is reminiscent of the well-known
seesaw mechanism used to explain the smallness of neutrino masses \cite{neutrino}.
One way to obtain this result is to have the density of states scale as
$\rho(p^2)\sim p^4/\muP^4$, and place the UV cutoff at $\muSUSY$, beyond which the 
bosonic and fermionic contributions cancel. 
This would yield $\epsilon^4 \sim \muSUSY^8/\muP^4$.
Unfortunately, this density of states is problematic since $p^4/\muP^4 \ll 1$
for the entire integration region, so we are effectively suppressing everything.
Furthermore, to obtain this suppression, we must have
$A(p^2)\sim (\muP/p)^{4/3} \gg 1$, making the effective value of $\hbar$, 
and thus the size of the quantum cell, huge at low energies in clear contradiction to reality.

In retrospect, this result is not surprising since raising the UV cutoff from
$\sqrt{\muP\mu_0}\sim 10^{-3}\,\mathrm{eV}/c$ to much higher values naturally requires
the drastic suppression of contributions from below the cutoff.
Thus, it is clear that the modification to the density of states, as suggested by the MLUR,
by itself cannot solve the cosmological constant problem.

\subsection{Need for a UV/IR relation and a Dynamical Energy-Momentum Space}

In the above discussion of summing over momentum states, the unstated assumption was that
states at different momentum scales were independent, and that their total effect on the
vacuum energy was the simple sum of their individual contributions.
Of course, this assumption is the basis of the decoupling between small (IR) and large (UV) 
momentum scales which underlies our use of effective field theories.
However, there are hints that this assumption is what needs to be reevaluated
in order to solve the cosmological constant problem.

First and foremost, the expression for the vacuum energy density 
itself, $H_0^2 c^2/\GN = \hbar c/\ellP^2\ell_0^2$, is dependent upon an IR scale $\ell_0$ and
a UV scale $\ellP$, suggesting that whatever theory that explains its value
must be aware of both scales, and have some type of dynamical connection between them.
Note that effective QFT's are not of this type but string theory is, given 
the UV/IR mixing relations discovered in several contexts as mentioned in the introduction.

Second, the contributions of the sub-Planckian modes ($p<\muP$) independently
by themselves are clearly too large, and there is a limit to the tweaking that can be 
done to the density of states in the IR since those modes undeniably exist.
The only way out of the dilemma would be to cancel the contribution of the
IR sub-Planckian modes against those of something else, say that coming from the
UV trans-Planckian modes ($p>\muP$) by introducing a dynamical connection between the two regimes \cite{Banks:2001zj}.  

That the sub-Planckian ($p<\muP$) and trans-Planckian ($p>\muP$) modes should cancel against each other is also
suggested by the following argument:
Consider how the MLUR, Eq.~(\ref{MLUR2}), would be realized in field theory.
The usual Heisenberg relation $\delta x\,\delta p = \hbar/2$ is a simple consequence of the fact that
coordinate and momentum spaces are Fourier transforms of each other.  The more one wishes to 
localize a wave-packet in coordinate space (smaller $\delta x$), 
the more momentum states one must superimpose (larger $\delta p$).
In the usual case, there is no lower bound to $\delta x$: one may localize the wave-packet as
much as one likes by simply superimposing states with ever larger momentum, and thus ever shorter wavelength, 
to cancel out the tails of the coordinate space distributions.  On the other hand, 
the MLUR implies that if one keeps on superimposing states with momenta beyond $\muP = 1/\sqrt{\beta}$,
then $\delta x$ ceases to decrease and starts increasing instead.
(See Fig~\ref{MLURfig}.) 
The natural interpretation of such a phenomenon would be that the trans-Planckian modes 
($p>\muP$) when superimposed with the sub-Planckian ones ($p<\muP$)
would `jam' the sub-Planckian modes and prevent them from canceling out the tails of the wave-packets 
effectively.
The mechanism we envision here is analogous to the `jamming' behavior seen
in non-equilibrium statistical physics, in which systems 
are found to freeze with increasing temperature \cite{jam}.
In fact, it has been argued
that such ``freezing by heating'' could be characteristic of a background
independent quantum theory of gravity \cite{chia}.

We should also note, that in our calculation presented above, 
the phase space over which the integration was performed was fixed and flat.  
Quantum gravity will naturally change the situation, leading to
a fluctuating dynamical spacetime background.
Furthermore, the MLUR implies that energy-momentum space will be a fluctuating
dynamical entity as well \cite{CurvedMomentumSpace,Chang:2010ir}.
First, the necessity of ``jamming'' between the sub-Planckian and trans-Planckian modes
to implement the MLUR in field theory
clearly illustrates that momentum space cannot be the simple Fourier transform of coordinate space,
but must rather be an independent entity.\footnote{%
Introducing a momentum space independent from coordinate space in field theory would make
the wave-particle duality more complete in a sense, since for particles, momenta and coordinates
are independent until the equation of motion is imposed \cite{Bars:2010xi}.}
Second, the quantum properties of spacetime geometry may be 
understood in terms of effective expressions that involve the spacetime uncertainties:
\begin{equation}
g_{ab}(x)\, d x^a d x^b  \quad\to\quad g_{ab}(x)\, \delta x^a \delta x^b\;.
\end{equation}
The UV/IR relation $\delta x\sim \hbar\beta\,\delta p$ in the trans-Planckian region 
implies that this geometry of spacetime uncertainties can be transferred directly to 
the space of energy-momentum uncertainties, endowing it with a geometry as well:
\begin{equation}
g_{ab}(x)\, \delta x^a \delta x^b \quad\to\quad G_{ab}(p)\, \delta p^a \delta p^b\;.
\end{equation}
The usual intuition that local properties in spacetime
correspond to non-local features of energy-momentum space
(as implied by the canonical uncertainty relations) is
obviated by the linear relation between the uncertainties in coordinate space
and momentum space.

What would a dynamical energy-momentum space entail? 
Let us speculate.
It has been argued that a dynamical spacetime,
with its foamy UV structure \cite{Wheeler:1957mu}, would manifest itself in the IR
via the uncertainties in the measurements of 
global spacetime distances as \cite{Wigner:1957ep,ng,AmelinoCamelia:1994vs,Diosi:1989hy}:
\begin{equation}
\delta\ell \;\sim\; \sqrt{\ell\,\ellP}\;,
\end{equation}
a relation which is reminiscent of the famous result for Brownian motion derived by Einstein \cite{EinsteinBrown},
and is also covariant in $3+1$ dimensions.
Let us assume that a similar `Brownian' relation holds in energy-momentum space
due to its `foaminess' \cite{Chang:2010ir}:
\begin{equation}
\delta \mu \sim \sqrt{\mu\,\muP}\;.
\end{equation}
If the energy-momentum space has a finite size, a natural UV cutoff, at $\mu_+ \gg \muP$,\footnote{%
A maximum energy/momentum is introduced in deformed special relativity \cite{Magueijo:2001cr}.}
then its fluctuation $\delta\mu_+$ will be given by 
$\delta\mu_+ = \sqrt{\mu_+\,\muP} \gg \muP$.
The MLUR implies that the mode at this scale must cancel, or `jam,' against another
which shares the same $\delta x$, namely, the mode with an uncertainty given by
$\delta\mu_- = \muP^2/\delta\mu_+ = \muP\sqrt{\muP/\mu_+} = \sqrt{\mu_-\muP} \ll \muP$, 
that is:
\begin{equation}
\mu_- \;=\; \dfrac{\muP^2}{\mu_+} \;=\; \dfrac{\delta\mu_-^2}{\muP} \;\ll\; \muP\;.
\end{equation}
All modes between $\mu_-$ and $\mu_+$ will `jam.'
Therefore, $\mu_-$ will be the effective UV cutoff of the momentum integral and not $\mu_+$,
which would yield
\begin{equation}
{\epsilon}^4 \;\sim\; \mu_{-}^4 
\;\sim\; \dfrac{\delta \mu_{-}^8}{\muP^4}
\;\sim\; \dfrac{\muP^8}{\mu_+^4}\;.
\end{equation}
This reproduces the seesaw formula, Eq.~(\ref{seesaw}),
and if $\delta\mu_-\sim \mbox{few TeV}/c$, we obtain the correct
cosmological constant.

\section{Outlook: What is string theory?}

In the concluding section we wish to discuss a few implications of our work for non-perturbative string theory and the question: What is string theory \cite{joewhat}?  Our discussion of this difficult 
question, being limited by the scope of our work on the minimal length, will neccesarily be a bit speculative.

Our toy model for the MLUR was essentially algebraic.
As such, it raises the possibility that more general algebraic structures may play a key role 
in non-perturbative string theory.  
In the introductory section, we mentioned that the MLUR 
is motivated by the scattering of string like excitations in first quantized string theory.
If one takes into account other objects in non-perturbative string theory, such as $D$-branes, 
one is lead to the STUR, Eq.~(\ref{STUR}), proposed by Yoneya.
The STUR generalizes the MLUR, and
can be further generalized to a cubic form (motivated by M-theory) \cite{yoneya}
\begin{equation}
\delta x\,\delta y\,\delta t \;\sim\; \ellP^3/c\;.
\end{equation}
Given the usual interpretation of the canonical Heisenberg uncertainty relations
in terms of fundamental commutators, one might look for
the associated cubic algebraic structures in string theory.

Another hint of cubic algebraic structure
appears in the non-perturbative
formulation of open string field theory by Witten \cite{Witten:1985cc}.
The Witten action for the classical open string field, $\Phi$, is of an abstract Chern-Simons type
\begin{equation}
S_\mathrm{o}(\Phi) \;=\; \int \Phi \star \Phi \star \Phi\;.
\end{equation}
Here the star product is
defined by the world-sheet path integral,
\begin{equation}
F \star G \;=\; \int DX\;F(X)\,G(X)\;\exp\!\left[\frac{i}{\alpha'}\,S_\mathrm{P}(X)\right]\;,
\end{equation}
which is in turn determined by the world-sheet Polyakov action
\begin{equation}
S_\mathrm{P} (X) 
\;=\; \frac{1}{2} \int d^2 \sigma\,\sqrt{-g}\,g^{ab}\,\partial_a X^i\,\partial_b X^j\;G_{ij} + \cdots \;.
\end{equation} 
The fully quantum open string field theory is then, in principle, defined by yet another
path integral in the infinite dimensional space of $\Phi$, i.e. 
\begin{equation}
\int D\Phi\;\exp\!\left[\dfrac{i}{g_c}\,S_\mathrm{o}(\Phi)\right]\;.
\end{equation}
A more general, and in principle non-associative structure, appears in Strominger's formulation
of closed string field theory, which is also cubic \cite{Strominger:1987ad}. 
Strominger's paper 
mentions the relevance of the 3-cocycle structure for this formulation of closed string field theory.
Very schematically
\begin{equation}
S_\mathrm{c}(\Psi) \;=\; \int \Psi \times (\Psi \times \Psi)\;,
\end{equation}
where $\times$ is a non-associative product defined in Ref.~\cite{Strominger:1987ad}.
(For the role of non-associativity in the theories of gravity,
and a relation between Einstein's gravity and non-associative
Chern-Simons theory, see \cite{okubo}.)

Is there an underlying algebraic structure that could give rise to these cubic structures? In our toy model the 2-bracket appears quite naturally.
Such structures can be naturally generalized into 3-brackets.
For example, the usual Lie algebra structure known from gauge theories,
$\bigl[\,T_A,\,T_B\,\bigr] = f_{ABC}\,T_{C}$,
where the structure constants $f_{ABC}$ satisfy the usual Jacobi identity, 
seems to be naturally generalized to a triple algebraic structure
\begin{equation}
\bigl[\,T_A,\,T_B,\,T_C\,\bigr] \;=\; f_{ABCD}\,T_{D}\;,
\end{equation}
where
\begin{equation}
\bigl[\,A_i,\,A_j,\,A_k\,\bigr] \;\equiv\; \epsilon^{abc}\,A_a\,A_b\,A_c\;,
\label{TripleBracketDef}
\end{equation}
with the structure constants $f_{ABCD}$ satisfying a quartic fundamental identity \cite{Awata:1999dz,Nambu:1973qe}.
These structures occur in the context of the theory of $N$-membranes \cite{vector}. 
They are also present in more elementary examples.
Consider a charged particle $e$ of mass $m$ in the external magnetic field $\mathbf{B}$.
As is well known, the velocities $\vhat_a$
satisfy the commutation relation 
\begin{equation}
\bigl[\,\vhat_i,\,\vhat_j\,\bigr] \;=\; i\,\dfrac{e \hbar}{m^2}\,\epsilon_{ijk}\,B_k\;,
\end{equation}
as well as the triple commutation relation, the associator, given by \cite{jackiw}
\begin{equation}
 \bigl[\,\vhat_1,\bigl[\,\vhat_2,\,\vhat_3\,\bigr]\bigr]
+\bigl[\,\vhat_2,\bigl[\,\vhat_3,\,\vhat_1\,\bigr]\bigr]
+\bigl[\,\vhat_3,\bigl[\,\vhat_1,\,\vhat_2\,\bigr]\bigr]
\;=\; \dfrac{e \hbar^2}{m^3}\,\partial_i B_i
\;.
\end{equation}
This associator is zero, and thus trivial,
in the absence of magnetic monopoles: $\partial_i B_i =0$.
Note that the triple bracket defined in Eq.~(\ref{TripleBracketDef})
is ``one-half'' of the
associator since
\begin{eqnarray}
\bigl[\,A,\widehat{B,\,C}\bigr] 
& \equiv & \epsilon^{abc}\,A_a(A_b\,A_c)
\;=\;
 A\bigl[\,B,\,C\,\bigr] 
+B\bigl[\,C,\,A\,\bigr]
+C\bigl[\,A,\,B\,\bigr]
\;,\cr
\bigl[\,\widehat{A,\,B},\,C\bigr] 
& \equiv & \epsilon^{abc}\,(A_a\,A_b)A_c
\;=\;
 \bigl[\,B,\,C\,\bigr]A
+\bigl[\,C,\,A\,\bigr]B
+\bigl[\,A,\,B\,\bigr]C
\;.
\end{eqnarray}
The presence of monopoles is an indicator of a 3-cocycle \cite{jackiw}.
The triple commutator has also been encountered in the study of of closed string dynamics \cite{Blumenhagen:2010hj}.

What would be the role of such a general algebraic structure for the foundations of
string theory? Given the general open-closed string relation (the closed strings being in some
sense the bounds states of open strings) the non-commutative and non-associative 
algebraic structures might be related as 
in some very general and abstract form of the celebrated AdS/CFT duality \cite{adscft}.
We recall that in the AdS/CFT correspondence, one computes the on-shell
bulk action $S_\mathrm{bulk}$ and relates it to the appropriate boundary
correlators. The conjecture is that
the generating functional of the vacuum correlators of
the operator $\Ohat$ for
a $d$-dimensional conformal field theory (CFT) is given by the
partition function $Z(\phi)$ in (Anti-de-Sitter) AdS${}_{d+1}$ space:
\begin{equation}
\left\langle \exp\left(\int\!J\Ohat \right)\right\rangle \;=\; Z (\phi) 
\qquad\to\qquad
\exp\Bigl[\,-S_\mathrm{bulk}(\,g,\,\phi,\,\cdots)\,\Bigr]
\;,
\end{equation}
where in the semiclassical limit the
partition function becomes $Z= \exp(-S_\mathrm{bulk})$.
Here $g$ denotes the metric of the AdS${}_{d+1}$ space,
and the boundary values of the bulk field, $\phi$, are given by
the sources, $J$, of the boundary CFT.
Essentially, one reinterprets the RG flow of the
boundary non-gravitational theory in terms of
bulk gravitational equations of motion, and then
rewrites the generating functional of vacuum correlators
of the boundary theory in terms of a semi-classical
wave function of the bulk ``universe'' with specific boundary
conditions.

In view of our comments on the general algebraic structures in string theory,
it is tempting to propose an extension of this duality in a more abstract sense
of open and closed string field theory, and the relationship between
the non-commutative and non-associative structures
\begin{equation}
\left\langle \exp\left(\int\! J\Ohat(\Phi) \right)\right\rangle_\mathrm{o} 
\;=\; Z (\Psi) 
\qquad\to\qquad
\exp\Bigl[\,- S_{c}(\Psi)\,\Bigr]
\;.
\end{equation}
The ``boundary'' in this abstract case has to be defined algebraically, as a region
of the closed string Hilbert space on which the 3-cocycle anomaly vanishes.
Inside the region, the 3-cocycle would be non-zero. In this way, we would have
more abstract definitions of the ``boundary'' and ``bulk.''
In some sense, this relation would look like a generalized Laplace transform of an exponental
of a cubic
expression giving another exponential of a cubic expression, as with the asymptotics of
the Airy function $\int dx \exp(tx - x^3) \sim \exp(-t^3/2)$.

Finally, following our discussion of the vacuum energy problem in the previous section,
it seems natural that any more fundamental formulation of 
string theory would have to work on a curved momentum space.
This would mesh nicely with the ideas presented in Refs.~\cite{chia} and \cite{CurvedMomentumSpace}.
If curved energy-momentum space is
crucial in quantum gravity (and thus string theory)
for the solution of the vacuum energy problem, then we are
naturally lead to question the usual formulation of string theory as a 
canonical quantum theory.
Also, if the vacuum energy can be made small, what physical principle selects such
a vacuum? This leads to the question of background independence and vacuum selection.
The issue of background independence in string theory is
 that the fundamental equations should not select a quantum theory the same way Einstein's
gravitational equations do not select any geometry; only asymptotic or symmetry
conditions select a geometry. Again, we are back to the questions regarding the role of
general quantum theory in the most fundamental formulation of
string theory. Note that such discussion of general quantum theory
also sheds light on the question of time evolution and the problem
of time in string theory \cite{chia}.

\section*{Acknowledgments}

We wish to thank Yang-Hiu He for inviting us to write this review.
DM, TT, and ZL are supported in part by
the U.S. Department of Energy, grant DE-FG05-92ER40677, task A.


\end{document}